\documentclass[notitlepage,nofootinbib,preprintnumbers,amssymb,superscriptaddress]{revtex4-1}

\usepackage{amsfonts,amssymb,amsmath,graphicx,color,bm}
\definecolor{ultramarine}{rgb}{0.07, 0.04, 0.56}
\definecolor{cadmiumgreen}{rgb}{0.0, 0.42, 0.24}
\definecolor{indigo(dye)}{rgb}{0.0, 0.25, 0.42}
\usepackage[dvipdfmx]{hyperref}
\hypersetup{
colorlinks=true,
citecolor=ultramarine,
linkcolor=cadmiumgreen,
urlcolor=indigo(dye),
}

\newcommand{\fr}[2]{\frac{#1}{#2}}
\newcommand{\pa}{\partial}
\newcommand{\ti}{\tilde}
\newcommand{\na}{\nabla}

\newcommand{\bra}[1]{\left( #1 \right)}
\newcommand{\brb}[1]{\left[ #1 \right]}
\newcommand{\brc}[1]{\left\{ #1 \right\}}
\newcommand{\pb}[1]{\brc{#1}_{\rm P}}
\newcommand{\be}{\begin{equation}}
\newcommand{\ee}{\end{equation}}
\newcommand{\bem}{\begin{bmatrix}}
\newcommand{\eem}{\end{bmatrix}}
\newcommand{\al}{\alpha}
\newcommand{\ga}{\gamma}
\newcommand{\e}{\epsilon}
\newcommand{\ka}{\kappa}
\newcommand{\La}{\Lambda}
\newcommand{\la}{\lambda}
\newcommand{\si}{\sigma}
\newcommand{\vp}{\varphi}
\newcommand{\Om}{\Omega}
\newcommand{\mn}{{\mu \nu}}
\newcommand{\mA}{\mathcal{A}}
\newcommand{\mB}{\mathcal{B}}
\newcommand{\mC}{\mathcal{C}}

\newcommand{\mE}{\mathcal{E}}
\newcommand{\mF}{\mathcal{F}}
\newcommand{\mG}{\mathcal{G}}
\newcommand{\mH}{\mathcal{H}}
\newcommand{\mK}{\mathcal{K}}
\newcommand{\mM}{\mathcal{M}}
\newcommand{\mN}{\mathcal{N}}
\newcommand{\mP}{\mathcal{P}}
\newcommand{\mR}{\mathcal{R}}
\newcommand{\mW}{\mathcal{W}}
\newcommand{\HL}{H_{\rm L}}
\newcommand{\HS}{H_{\rm S}}

\begin{document}

\preprint{RESCEU-9/17, RUP-17-15}

\title{Extended mimetic gravity: Hamiltonian analysis and gradient instabilities}

\author{Kazufumi Takahashi}
\affiliation{Research Center for the Early Universe (RESCEU), Graduate School of Science, The University of Tokyo, Tokyo 113-0033, Japan}
\affiliation{Department of Physics, Graduate School of Science, The University of Tokyo, Tokyo 113-0033, Japan}
\author{Tsutomu Kobayashi}
\affiliation{Department of Physics, Rikkyo University, Toshima, Tokyo 171-8501, Japan}

\begin{abstract}
We propose a novel class of degenerate higher-order scalar-tensor theories as an extension of mimetic gravity.
By performing a noninvertible conformal transformation on ``seed'' scalar-tensor theories which may be nondegenerate, we can generate a large class of theories with at most three physical degrees of freedom.
We identify a general seed theory for which this is possible.
Cosmological perturbations in these extended mimetic theories are also studied.
It is shown that either of tensor or scalar perturbations is plagued with gradient instabilities, except for a special case where the scalar perturbations are presumably strongly coupled, or otherwise there appear ghost instabilities.
\end{abstract}

\maketitle

\section{Introduction}\label{sec:intro}

When one constructs a field theory with higher derivatives, a guiding principle comes from the theorem of Ostrogradsky~\cite{Woodard:2015zca}, which states that any theory described by a nondegenerate higher derivative Lagrangian suffers from instabilities of the Ostrogradsky ghost.
Here, a Lagrangian $L$ with higher-order derivatives is said to be nondegenerate if its kinetic matrix (i.e., the second derivatives of $L$ with respect to the velocities associated with the higher derivatives) is nondegenerate.
Therefore, a theory without the Ostrogradsky ghost, often referred to as a {\it healthy} theory, must have a degenerate Lagrangian.\footnote{Even if a theory circumvents the problem of Ostrogradsky ghost, there could be some other instabilities. In this sense, the ``healthiness'' is a necessary but not sufficient condition for a theory to be fully viable.}
This requirement poses tight constraints on the higher derivative structure of healthy field theories~\cite{Motohashi:2016ftl,Crisostomi:2017aim}.

Within (single-field) scalar-tensor theories in four dimensions, the Horndeski theory~\cite{Horndeski:1974wa} (or its equivalent formulation known as generalized Galileons~\cite{Deffayet:2011gz,Kobayashi:2011nu}) provides a basic ground for studying a wide class of such healthy theories having three degrees of freedom~(DOFs), since it is the most general theory that yields second-order Euler-Lagrange equations.
There are further possibilities of healthy theories beyond the Horndeski class, such as Gleyzes-Langlois-Piazza-Vernizzi (GLPV) theories~\cite{Gleyzes:2014dya} and quadratic/cubic degenerate higher-order scalar-tensor (DHOST) theories~\cite{Langlois:2015cwa,Crisostomi:2016czh,BenAchour:2016fzp}.
Those quadratic/cubic DHOST theories form the broadest class of healthy scalar-tensor theories known so far.\footnote{We restrict ourselves to theories that possess general covariance.
If one relaxes this requirement to only spatial covariance, yet broader classes of theories can be obtained such as extended Galileons~\cite{Gao:2014soa,Fujita:2015ymn}.
However, even though general covariance is broken apparently, it can always be restored by introducing a St\"{u}ckelberg field.
Then, it turns out that those theories generically yield Ostrogradsky ghosts.
See Refs.~\cite{Blas:2009yd,Langlois:2015cwa} for examples of such theories.
The apparent healthy nature of extended Galileons is a consequence of the unitary gauge~$\phi=t$ chosen for the construction of the theory.
This misleading gauge choice could eliminate higher derivatives from a given action and change the DOFs~\cite{Blas:2009yd,Deffayet:2015qwa,Langlois:2015cwa,Langlois:2015skt,Crisostomi:2016tcp}, though it is a complete gauge fixing~\cite{Motohashi:2016prk}.}
However, these theories are obtained under the assumption that the Lagrangian depends on up to quadratic/cubic order in $\na_\mu\na_\nu\phi$ (hence the name ``quadratic/cubic DHOST''), and thus the very boundary of healthy scalar-tensor theories remains unknown.
To go even further, disformal transformations~\cite{Bekenstein:1992pj} may play a key role:
Some quadratic/cubic DHOST theories are obtained from the Horndeski theory via disformal transformation~\cite{BenAchour:2016fzp}, which motivates us to think that the quadratic/cubic DHOST class could further be extended in the same manner.
Unfortunately, this is not the case because the set of cubic scalar-tensor theories is closed under disformal transformations.
Therefore, it seems difficult to generate a new class of healthy theories starting from known healthy theories.

In light of this situation, it is natural to ask whether healthy theories can be generated from {\it nondegenerate} theories that contain extra ghost DOFs.
This cannot be achieved by an invertible transformation (i.e., a field redefinition such that there is one-to-one correspondence between the old and new sets of field variables) since it does not change the number of physical DOFs~\cite{Deruelle:2014zza,Domenech:2015tca,Takahashi:2017zgr}.
However, the possibility is still open for {\it noninvertible} transformations.
In this context, an interesting theory is {\it mimetic gravity}~\cite{Chamseddine:2013kea} (see Ref.~\cite{Sebastiani:2016ras} for a review).
This theory is obtained from the Einstein-Hilbert action of general relativity (GR) by performing the following noninvertible conformal transformation:
	\be
	\ti{g}_\mn=-Xg_\mn, \quad X\equiv g^\mn\pa_\mu\phi\pa_\nu\phi,\label{mimtrnsfI}
	\ee
where $\ti{g}_\mn$ and $g_\mn$ denote respectively the metrics of the original frame (namely, the GR frame) and the new frame.
This theory can mimic the behavior of pressureless dust in GR~\cite{Chamseddine:2013kea}, and thus is one of the candidates for dark matter.
The above formulation of mimetic gravity can be straightforwardly extended by generalizing the ``seed'' action of the original frame to that of a scalar-tensor theory such as the Horndeski theory~\cite{Arroja:2015wpa}.
The noninvertibility of the transformation~\eqref{mimtrnsfI} is manifested by the fact that the right-hand side of Eq.~\eqref{mimtrnsfI} is invariant under the conformal transformation~$g_\mn\to \Om^2g_\mn$, with $\Om$ being an arbitrary function of spacetime.
Although the resultant theory written in terms of $(g_\mn,\phi)$ has higher-order derivatives of $\phi$, the authors of Ref.~\cite{Chaichian:2014qba} performed a Hamiltonian analysis to show that the theory has only three DOFs due to the local conformal symmetry introduced by the transformation~\eqref{mimtrnsfI}.
Remarkably, theories with 3 DOFs could be obtained even if one starts from a large class of nondegenerate higher-order scalar-tensor theories instead of GR or healthy scalar-tensor theories with 3 DOFs.
Indeed, the author of Ref.~\cite{Kluson:2017iem} performed a Hamiltonian analysis of the mimetic theory resulting from
	\be
	S=\int d^4x\sqrt{-g}\brb{\fr{1}{2}\mR+f(\Box\phi)}, \label{R+f(boxphi)}
	\ee
with $\mR$ being the 4-dimensional Ricci scalar and $f$ an arbitrary scalar function, and proved that the model has at most three DOFs.\footnote{Strictly speaking, the author of Ref.~\cite{Kluson:2017iem} used an alternative formulation of mimetic gravity proposed in Ref.~\cite{Golovnev:2013jxa} (see \S \ref{ssec:remark}).}
(Note here that it might be more appropriate to use the notation like $\ti{\mR}$ etc.~to write the seed action, but in this paper we do not do so in order to avoid too many tildes.
Therefore, to generate a mimetic theory one should replace $g_{\mu\nu}$ in the seed action by $-Xg_{\mu\nu}$.)
In the fluid description, such higher derivative terms typically introduces imperfectness~\cite{Chamseddine:2014vna} and the scalar DOF acquires a nonzero sound speed~\cite{Mirzagholi:2014ifa}, which may solve some of the small-scale problems like the missing-satellite problem and the core-cusp problem~\cite{Capela:2014xta}.

In the present paper, we consider generic scalar-tensor theories that could possess an unwanted extra DOF as a generalization of Eq.~\eqref{R+f(boxphi)}, and perform the noninvertible transformation~\eqref{mimtrnsfI} on them, as suggested in Ref.~\cite{Liu:2017puc}.
We show explicitly that the extended mimetic gravity models obtained thus have at most three DOFs based on Hamiltonian analysis.
This turns out to be true also for models obtained via noninvertible disformal transformation that is more general than~\eqref{mimtrnsfI}.
Due to the diversity of the original theories, many of the mimetic theories lie outside the quadratic/cubic DHOST class and they cannot be obtained by (invertible) disformal transformation of any known class of healthy scalar-tensor theories.
Nevertheless, such mimetic theories have a problem on cosmological perturbations:
It was demonstrated in Refs.~\cite{Ijjas:2016pad,Firouzjahi:2017txv} that the mimetic model obtained from the action of the form~\eqref{R+f(boxphi)} has ghost/gradient instabilities in cosmological perturbations.
The simplest version of mimetic gravity is closely related to the low-energy limit of Ho\v{r}ava-Lifshitz gravity~\cite{Horava:2009uw,Ramazanov:2016xhp}, and the same instability was also pointed out in the latter context in Ref.~\cite{Sotiriou:2009bx}.
Concerning this point, the authors of Ref.~\cite{Hirano:2017zox} developed an effective theory of cosmological perturbations in mimetic theories, and showed that the instability can be cured by introducing nonminimal derivative couplings to gravity.
In this paper, we also study the linear stability of cosmological perturbations in our extended mimetic gravity and show that the models obtained in the above explained way generically exhibit the very same problem of ghost/gradient instabilities (except for the special case in which scalar perturbations appear to be strongly coupled).

This paper is organized as follows.
In \S \ref{sec:ADMST}, we begin with presenting the general seed action which we use to generate a variety of mimetic gravity theories, and then perform the Arnowitt-Deser-Misner (ADM) decomposition of the seed action.
Then, in \S \ref{sec:mim} we transform the seed action to its mimetic counterpart via the noninvertible conformal transformation~\eqref{mimtrnsfI}, and analyze the resultant theory in the language of the Hamiltonian formalism.
Cosmological perturbations in the extended mimetic gravity models are discussed in \S \ref{sec:pert}.
Finally, we draw our conclusions in \S \ref{sec:conc}.

\section{ADM representation of the seed scalar-tensor theory}\label{sec:ADMST}

\subsection{The seed action}

We start from the following general action:
	\be
	S=\int d^4x\sqrt{-g}\brb{f_2\mR+f_3\mG^\mn \na_\mu\na_\nu\phi+F(g_\mn,\phi,\na_\mu\phi,\na_\mu\na_\nu\phi)}, \label{genac}
	\ee
where $\mG_\mn$ is the 4-dimensional Einstein tensor, $f_2$ and $f_3$ are arbitrary functions of $(\phi,X)$, and $F$ denotes any scalar quantity constructed from the metric, the scalar field, and its derivatives up to second order.

Using the action of the form~\eqref{genac} as a seed, we perform the conformal transformation mentioned in \S \ref{sec:intro}:
	\be
	g_{\mu\nu}~\to~\ti{g}_\mn=-Xg_\mn, \label{mimtrnsf}
	\ee
where $\ti{g}_\mn$ is identified as the metric in the original frame~\eqref{genac}, while $g_\mn$ is now the metric of the new theory.
This transformation is noninvertible as the right-hand side is invariant under conformal transformation of $g_\mn$.
As a result, the new theory acquires a local conformal symmetry.
See the recent paper by Horndeski~\cite{Horndeski:2017rtl} for conformally invariant scalar-tensor theories.
We comment on the relation between the
Horndeski's conformally invariant theory and our extended mimetic gravity in Appendix~\ref{app:Horndeski}.
In general, one could consider noninvertible disformal transformations other than Eq.~\eqref{mimtrnsf}.
However, as far as our purpose is concerned, we lose no generality by restricting ourselves to the transformation~\eqref{mimtrnsf} since any noninvertible disformal transformation can be recast in this form.
This point is addressed in Appendix~\ref{app:disf}.

In Eq.~\eqref{genac} we consider only particular couplings between the scalar field and the curvature tensors.
Clearly, they are the same as those found in the Horndeski theory~\cite{Horndeski:1974wa}.
Other types of couplings such as $\mR(\Box\phi)^2$, $f(\phi,X)\Box \mR$, $\cdots$ would give rise to unwanted extra DOFs in the resultant mimetic theory.
This point will become clear in the Hamiltonian analysis below:
If one considers the couplings other than the first two terms in Eq.~\eqref{genac}, then one would be forced to introduce some new velocities, leading to the extra DOFs (see \S \ref{ssec:remark}).
For the same reason, we do not include third or higher derivatives of $\phi$ in Eq.~\eqref{genac}.

The class of theories defined by Eq.~\eqref{genac} includes all the known healthy scalar-tensor theories that possess general covariance, such as the Horndeski theory~\cite{Horndeski:1974wa}, GLPV theories~\cite{Gleyzes:2014dya}, and quadratic/cubic DHOST theories~\cite{Langlois:2015cwa,Crisostomi:2016czh,BenAchour:2016fzp}.
Such healthy theories correspond to specific choices of the functions~$f_2$, $f_3$, and $F$.
However, for generic functions, the seed theory~\eqref{genac} in its original frame would have Ostrogradsky ghosts.
Nevertheless, as we will show, the noninvertible transformation~\eqref{mimtrnsf} makes the resultant theory degenerate, leaving only 3 DOFs.


\subsection{ADM decomposition}

To proceed to a Hamiltonian analysis, we first express the seed action~\eqref{genac} in the ADM form and then perform the conformal transformation~\eqref{mimtrnsf} written in terms of the ADM variables.
In this subsection we present some technical detail for recasting Eq.~\eqref{genac} into the ADM form.

The metric in the ADM form is given by
	\be
	ds^2=g_\mn dx^\mu dx^\nu=-N^2dt^2+\ga_{ij}(dx^i+N^idt)(dx^j+N^jdt).
	\ee
The spatial metric~$\ga_{ij}$ is used to raise and lower spatial indices~$i,j,\cdots$.
The unit normal vector to a constant-time hypersurface is denoted as $n_\mu\equiv -N\delta^0_\mu$, and the projection tensor as $h_\mn\equiv g_\mn+n_\mu n_\nu$.
The extrinsic curvature is defined by
	\be
	K_{ij}\equiv \fr{1}{2N}\bra{\dot{\ga}_{ij}-2D_{(i}N_{j)}},
	\ee
where $D_i$ is the covariant derivative with respect to $\ga_{ij}$.

We also introduce the following variables associated with time derivatives of the scalar field as in Ref.~\cite{Langlois:2015skt}:
	\begin{align}
	 A_\ast&\equiv n^\mu \na_\mu\phi=\fr{\dot{\phi}-N^iD_i\phi}{N}, \label{phiast} \\
	V_\ast&\equiv n^\mu n^\nu \na_\mu \na_\nu\phi=\fr{\dot{A}_\ast-D^i\phi D_iN-N^iD_i A_\ast}{N}. \label{Vast}
	\end{align}
Below, we will introduce a Lagrange multiplier and regard $A_\ast$ as an auxiliary variable which satisfies Eq.~\eqref{phiast} dynamically so that second-order time derivatives do not appear explicitly in the action.
Then, $V_\ast$ plays the role of the velocity of $A_\ast$.
Using $A_\ast$, the scalar kinetic term $X$ is written as
		\be
		X=- A_\ast^2+D^i\phi D_i\phi. \label{Xaux}
		\ee
It is worth emphasizing here that the unitary gauge $\phi=t$ is not imposed from the beginning since it would be misleading in some cases~\cite{Langlois:2015cwa,Langlois:2015skt}.
Nevertheless, in the case of mimetic gravity, it turns out that the number of physical DOFs is not changed by the unitary gauge fixing, as we will see in \S \ref{ssec:remark}.

With these notations, one can decompose $\na_\mu\phi$ and $\na_\mu \na_\nu\phi$ as
	\begin{align}
	\na_\mu\phi&=h_\mu^iD_i\phi-n_\mu A_\ast, \\
	\na_\mu \na_\nu\phi&=h_{(\mu}^ih_{\nu)}^j(D_iD_j\phi- A_\ast K_{ij})-2h_{(\mu}^in_{\nu)}(D_i A_\ast-K_{ij}D^j\phi)+n_\mu n_\nu V_\ast. \label{ddphi}
	\end{align}
These decompositions allow us to recast the third term of the action~\eqref{genac} in the form
	\be
	\int d^4x\sqrt{-g}F(g_\mn,\phi,\na_\mu\phi,\na_\mu \na_\nu\phi)=\int dtd^3x\brb{N\sqrt{\ga}L_F(\ga_{ij},\phi, A_\ast;K_{ij},V_\ast;D_i)+\La(N A_\ast+N^iD_i\phi-\dot{\phi})}, \label{scaac}
	\ee
where the concrete form of $L_F$ depends on that of $F$, and the last term with a Lagrange multiplier~$\Lambda$ fixes $ A_\ast$ so that it satisfies Eq.~\eqref{phiast}.

The first two terms in Eq.~\eqref{genac} involving the curvature tensors can be written in the ADM form by using the Gauss/Codazzi/Ricci equations.
To simplify the manipulation, we first perform integration by parts to move one of the derivative operators acting on $\phi$ to $f_3$.
Then, the result is given as follows:
	\begin{align}
	&\int d^4x\sqrt{-g}\bra{f_2\mR+f_3\mG^\mn \na_\mu\na_\nu\phi}=\int d^4x\sqrt{-g}\bra{f_2\mR-\mG^\mn \na_\mu\phi \na_\nu f_3} \nonumber \\
	&~~~=\int dtd^3xN\sqrt{\ga}\biggl\{ f_2\bra{R+K_{ij}^2-K^2}-2Kf_{2\perp}-2D^iD_if_2-\fr{1}{2}\bra{R-K_{ij}^2+K^2} A_\ast f_{3\perp} \nonumber \\
	&~~~~~~~~~~~~~~~~~~~~~~~~-\brb{R_{ij}-\fr{1}{2}\bra{R+K_{kl}^2-K^2}\ga_{ij}}D^i\phi D^jf_3+D_iD_j\bra{D^i\phi D^jf_3}-D_iD^i\bra{D_j\phi D^jf_3} \nonumber \\
	&~~~~~~~~~~~~~~~~~~~~~~~~+\bra{K\ga^{ij}-K^{ij}}\bra{2K_i^kD_k\phi D_jf_3+f_{3\perp} D_iD_j\phi+ A_\ast D_iD_j f_3}+\La(N A_\ast+N^iD_i\phi-\dot{\phi})\biggr\}, \label{curvac}
	\end{align}
where $R_{ij}$ and $R$ denote the 3-dimensional Ricci tensor and Ricci scalar, respectively.
Here, for a scalar function~$f(\phi,X)$ we have defined
	\be
	f_\perp(\ga_{ij},\phi, A_\ast;K_{ij},V_\ast;D_i)\equiv n^\mu\na_\mu f=f_\phi A_\ast-2f_X\bra{K_{ij}D^i\phi D^j\phi+ A_\ast V_\ast-D^i\phi D_i A_\ast}.
	\ee
Putting Eqs.~\eqref{scaac} and \eqref{curvac} together, one finds that the total action~\eqref{genac} can be written in the form
	\be
	S=\int dtd^3x\brb{N\sqrt{\ga}L_0(\ga_{ij},R_{ij},\phi, A_\ast;K_{ij},V_\ast;D_i)+\La(N A_\ast+N^iD_i\phi-\dot{\phi})}, \label{admac}
	\ee
where the dependence of $L_0$ on $N,N^i$ is encapsulated in $K_{ij}$ and $V_\ast$.
Note that in the actual expression of $L_0$ the spatial derivative $D_i$ does not act on $K_{ij}$ or $V_\ast$.

\section{Extended mimetic gravity}\label{sec:mim}

\subsection{Hamiltonian analysis}\label{ssec:mimHam}

In the previous section, we have written the seed action~\eqref{genac} in terms of the ADM variables to obtain Eq.~\eqref{admac}.
Now we move from the original frame~\eqref{admac} to another by performing the noninvertible conformal transformation~\eqref{mimtrnsf}, and thereby generate new mimetic gravity actions.

Under the transformation~\eqref{mimtrnsf}, the 3-dimensional quantities are transformed as follows:
	\begin{align}
	&\ti{N}=\sqrt{-X}N,~~~\ti{N}^i=N^i,~~~\ti{\ga}_{ij}=-X\ga_{ij},~~~\ti{A}_\ast=\fr{1}{\sqrt{-X}} A_\ast, \label{mimN} \\
	&\ti{R}_{ij}=R_{ij}+\fr{3}{4X^2}D_iXD_jX-\fr{1}{2X}D_iD_jX+\ga_{ij}\bra{\fr{1}{4X^2}D_kXD^kX-\fr{1}{2X}D_kD^kX}, \label{mimRij}\\
	&\ti{K}_{ij}=\sqrt{-X}\brb{K_{ij}-\fr{1}{X}\ga_{ij}\bra{K_{kl}D^k\phi D^l\phi + A_\ast V_\ast-D^k\phi D_k A_\ast}}, \label{mimKij} \\
	&\ti{V}_\ast=-\fr{1}{X^2}\bra{ A_\ast K_{ij}D^i\phi D^j\phi+V_\ast D^i\phi D_i\phi-D^i\phi D^j\phi D_iD_j\phi}, \label{mimVast}
	\end{align}
where original-frame variables are now denoted with tildes.
Note that the original-frame scalar kinetic term $\ti{X}$ is mapped to a constant:
	\be
	\ti{X}=\ti{g}^\mn\ti{\na}_\mu\phi\ti{\na}_\nu\phi=-\fr{1}{X}g^\mn\pa_\mu\phi\pa_\nu\phi=-1, \label{mimcon}
	\ee
which immediately leads to $\ti{\na}_\lambda\ti{X}=2\ti{g}^\mn\ti{\na}_\mu\phi\ti{\na}_\nu\ti{\na}_\la\phi=0$.
This means that any scalar quantity that contains a contraction of $\ti{\na}_\mu\phi$ and $\ti{\na}_\mu\ti{\na}_\nu\phi$ vanishes in the new frame.

It should be noted that in Eqs.~\eqref{mimKij} and~\eqref{mimVast} the velocity dependence appears only through the particular combination
	\be
	V_{ij}\equiv K_{ij}+\fr{V_\ast}{A_\ast}\ga_{ij}, \label{Vij}
	\ee
which is a consequence of the conformal symmetry introduced by performing the noninvertible conformal transformation~\eqref{mimtrnsf}.
In terms of this $V_{ij}$, Eqs.~\eqref{mimKij} and \eqref{mimVast} can be written as
	\begin{align}
	&\ti{K}_{ij}=\sqrt{-X}\brb{\bra{\delta^k_i\delta^l_j-\fr{D^k\phi D^l\phi}{X}\ga_{ij}}V_{kl}+\fr{D^k\phi D_k A_\ast}{X}\ga_{ij}}, \label{KijV1} \\
	&\ti{V}_\ast=-\fr{1}{X^2}D^i\phi D^j\phi \bra{A_\ast V_{ij}-D_iD_j\phi}.\label{VV1}
	\end{align}
As a side remark, the following identity holds,
	\be
	\ti{V}_{ij}=\sqrt{-X}V_{ij}-\bra{\fr{D^k\phi}{A_\ast}D_k\sqrt{-X}}\ga_{ij}, \label{mimVij}
	\ee
which will be used later.
Substituting Eqs.~\eqref{mimN},~\eqref{mimRij},~\eqref{KijV1}, and~\eqref{VV1} to the seed action~\eqref{admac}, we finally arrive at the action in the new frame:
	\be
	S=\int dtd^3x\brb{N\sqrt{\ga}L_{\rm M}(\ga_{ij},R_{ij},\phi, A_\ast;V_{ij};D_i)+\La(N A_\ast+N^iD_i\phi-\dot{\phi})}. \label{mimac0}
	\ee

When one is to switch to the Hamiltonian formalism, the nonlinear dependence on the velocity~$V_{ij}$ makes it difficult to express the action solely in terms of canonical coordinates and momenta.
To circumvent this technical issue, we introduce auxiliary variables~$B_{ij}$ with Lagrange multipliers~$\la^{ij}$ and rewrite the action as
	\begin{align}
	&S=S_{\rm M}[N,\ga_{ij},\phi, A_\ast,B_{ij}]+\int dtd^3x\brb{\La(N A_\ast+N^iD_i\phi-\dot{\phi})+N\la^{ij}(B_{ij}-V_{ij})}, \label{mimac} \\
	&S_{\rm M}[N,\ga_{ij},\phi, A_\ast,B_{ij}]\equiv \int dtd^3xN\sqrt{\ga}L_{\rm M}(\ga_{ij},R_{ij},\phi, A_\ast;B_{ij};D_i). \label{S0}
	\end{align}
Here, the lapse function~$N$ in front of $\la^{ij}$ was introduced so that the resultant Hamiltonian is linear in $N$ and $N^i$.
In the following, we perform a Hamiltonian analysis of the theory~\eqref{mimac}.
We define the canonical pairs as follows:
	\be
	\begin{pmatrix}
	N,&N^i,&\ga_{ij},&\phi,& A_\ast,&B_{ij},&\La,&\la^{ij}\\
	\pi_N,&\pi_i,&\pi^{ij},&p_\phi,&p_\ast,&p^{ij},&P,&P_{ij}
	\end{pmatrix}.
	\ee
These variables form a 50-dimensional phase space.

The canonical momenta are calculated from the action~\eqref{mimac} in the standard manner.
Since the action does not contain the velocities of $N,N^i,B_{ij},\La$, and $\la^{ij}$, the corresponding canonical momenta vanish:
	\be
	\pi_N=\pi_i=p^{ij}=P=P_{ij}=0,
	\ee
which provides primary constraints.
The canonical momenta for $\ga_{ij}$, $\phi$, and $A_\ast$ are given by
	\begin{align}
	\pi^{ij}&=\fr{\delta S}{\delta \dot{\ga}_{ij}}=-\fr{1}{2}\la^{ij}, \\
	p_\phi&=\fr{\delta S}{\delta \dot{\phi}}=-\La, \\
	p_\ast&=\fr{\delta S}{\delta \dot{A}_\ast}=-\fr{1}{ A_\ast}\ga_{ij}\la^{ij}.
	\end{align}
These expressions for the canonical momenta yield further primary constraints as
	\begin{align}
	\bar{\pi}^{ij}&\equiv \pi^{ij}+\fr{1}{2}\la^{ij}\approx 0, \\
	\bar{p}_\phi&\equiv p_\phi+\La\approx 0, \\
	\mC&\equiv A_\ast p_\ast-2\ga_{ij}\pi^{ij}\approx 0.
	\end{align}
It should be noted that $\mC$ is the generator of conformal transformation.
This relation between $\pi^{ij}$ and $p_\ast$ comes from Eq.~\eqref{Vij} with the identity
	\be
	 A_\ast\fr{\pa V_{ij}}{\pa\dot{A}_\ast}=2\ga_{kl}\fr{\pa V_{ij}}{\pa\dot{\ga}_{kl}}.
	\ee
To see the first-class nature of $\mC$, we construct a linear combination with $P_{ij}$ so that the resultant constraint weakly Poisson commutes with all the other primary constraints:
	\be
	\bar{\mC}\equiv \mC+2\la^{ij}P_{ij}= A_\ast p_\ast-2\ga_{ij}\pi^{ij}+2\la^{ij}P_{ij}.
	\ee
Note that the discussion so far does not depend on whether the original seed theory~\eqref{admac} is degenerate or not.
It might be possible that the resultant mimetic theory~\eqref{mimac} possesses additional primary constraints which have not been specified above, but it has nothing to do with the (non)degeneracy of the original theory.

Now the total Hamiltonian is obtained as
	\be
	H_T=\int d^3x\bra{N\mH+N^i\mH_i+\mu_N\pi_N+\mu^i\pi_i+\mu_{ij}\bar{\pi}^{ij}+u_\phi\bar{p}_\phi+u_\ast\bar{\mC}+u_{ij}p^{ij}+UP+U^{ij}P_{ij}},
	\ee
where
	\begin{align}
	\mH&\equiv -\sqrt{\ga}L_{\rm M}(\ga_{ij},R_{ij},\phi, A_\ast;B_{ij};D_i)+2\pi^{ij}B_{ij}+p_\phi A_\ast -\sqrt{\ga}D_i\bra{\fr{p_\ast}{\sqrt{\ga}}D^i\phi}, \\
	\mH_i&\equiv -2\sqrt{\ga}D^j\bra{\fr{\pi_{ij}}{\sqrt{\ga}}}+p_\phi D_i\phi+p_\ast D_i A_\ast +p^{jk}D_iB_{jk}-2\sqrt{\ga}D_j\bra{\fr{p^{jk}}{\sqrt{\ga}}B_{ik}}.
	\end{align}
Note that the last two terms in ${\cal H}_i$ are proportional to $p^{ij}$, which vanishes on the constraint surface.
Nevertheless, we keep these terms because they generate an infinitesimal spatial diffeomorphism of $B_{ij}$.

Let us calculate the time evolution of the primary constraints.
It is easy to see that the time evolution of $\bar{\pi}^{ij}$, $\bar{p}_\phi$, $P$, and $P_{ij}$ fixes the Lagrange multipliers~$U^{ij}$, $U$, $u_\phi$, and $\mu_{ij}$, respectively.
The time evolution of $\pi_N,\pi_i,p^{ij}$ leads to
	\begin{align}
	\dot{\pi}_N&=\pb{\pi_N,H_T}\approx -\mH, \\
	\dot{\pi}_i&=\pb{\pi_i,H_T}\approx -\mH_i, \\
	\dot{p}^{ij}&=\pb{p^{ij},H_T}\approx N\bra{\sqrt{\ga}\fr{\pa L_{\rm M}}{\pa B_{ij}}-2\pi^{ij}},
	\end{align}
where we have used the fact that the derivatives of $B_{ij}$ do not appear in $L_{\rm M}$.
Therefore, we find secondary constraints as
	\be
	\mH\approx 0,~~~\mH_i\approx 0,~~~\vp^{ij}\equiv \sqrt{\ga}\fr{\pa L_{\rm M}}{\pa B_{ij}}-2\pi^{ij}\approx 0.
	\ee
Note that the time evolution of $\bar{\mC}$ does not yield a new constraint because its Poisson bracket with the total Hamiltonian is written only by the constraints:
	\be
	\dot{\bar{\mC}}=\pb{\bar{\mC},H_T}=-N\mH+NB_{ij}\vp^{ij}+\pa_i\brb{N\fr{D^i\phi}{ A_\ast}\bra{\ga_{jk}\vp^{jk}-\mC}+N^i\mC}-2\mu_{ij}\bar{\pi}^{ij}+2U^{ij}P_{ij}\approx 0,
	\ee
where we have used the Noether identity with respect to the conformal symmetry~\eqref{conNoether}.

Having obtained all the secondary constraints, let us consider their time evolution.
Assuming that $L_{\rm M}$ depends at least quadratically on $B_{ij}$, the time evolution of $\vp^{ij}$ fixes the Lagrange multiplier $u_{ij}$.
To discuss the evolution of $\mH$ and $\mH_i$, we first note that $\mH\approx 0$ and $\mH_i\approx 0$ are expected to correspond to the Hamiltonian and momentum constraints, respectively.
Let us define smeared quantities as
	\be
	\HL[\mN]\equiv \int d^3x\mN\mH,~~~\HS[\mN^i]\equiv \int d^3x\mN^i\mH_i,
	\ee
where $\mN$ and $\mN^i$ are arbitrary test functions (and not the lapse function and the shift vector).
Then, with some manipulation their Poisson brackets are found to be the usual ones,
	\begin{align}
	\pb{\HS[\mN^i],\HS[\mM^i]}&=\HS[\mN^jD_j\mM^i-\mM^jD_j\mN^i], \label{pbSS} \\
	\pb{\HS[\mN^i],\HL[\mM]}&=\HL[\mN^iD_i\mM],
	\end{align}
where we have used the Noether identity associated with 3-dimensional diffeomorphism~\eqref{3diffNoether}.\footnote{When we apply the Noether identity~\eqref{3diffNoether}, we may replace the lapse function $N$ with a scalar test function $\mM$ in $\HL[\mM]$.}
The explicit calculation of $\pb{\HL[\mN],\HL[\mM]}$ is lengthy and tedious, and therefore we skip it.
Nevertheless, it is reasonable to conclude that $\mH\approx 0$ and $\mH_i\approx 0$ are the first-class constraints corresponding to general covariance and their time evolution does not yield any new constraints.
See Ref.~\cite{Langlois:2015skt} for the related discussion on this point.

To sum up, we have obtained the set of constraints as follows:
	\be
	\begin{array}{rl}
	9~{\rm first\verb|-|class}:&\pi_N,\pi_i,\bar{\mC},\mH,\mH_i, \\
	26~{\rm second\verb|-|class}:&\bar{\pi}^{ij},\bar{p}_\phi,p^{ij},P,P_{ij},\vp^{ij}.
	\end{array}
	\ee
These constraints reduce the phase-space dimension and
	\be
	\fr{1}{2}\bra{50-9\times 2-26}=3~{\rm DOFs}
	\ee
are left, which means that there is no unwanted extra DOF.
Note that this is the maximum possible number of physical DOFs that the theory~\eqref{mimac} has.
Even if the evolution of $\vp^{ij}$ yields some additional constraint as opposed to the above argument, it never increases the number of DOFs.
Note also that the original theory~\eqref{genac} could have a different number of physical DOFs.
The above result holds irrespective of whether we start from GR with two DOFs (as in original mimetic gravity~\cite{Chamseddine:2013kea}) or generic nondegenerate higher-order scalar-tensor theories with four DOFs.


\subsection{Some remarks}\label{ssec:remark}

Several remarks are in order.
First, we restricted ourselves to the case where the curvature tensors appear only in the form of $f_2(\phi,X)\mR$ and $f_3(\phi,X)\mG^\mn\na_\mu\na_\nu\phi$, since otherwise additional velocities must be introduced.
For example, in the case of mimetic $f(\mR)$ gravity~\cite{Nojiri:2014zqa}, one has to introduce a new velocity~$V_K\equiv n^\mu\na_\mu K$ in addition to $K_{ij}$ and $V_\ast$.
This results in an undesired extra DOF, which cannot be killed by the constraint corresponding to the conformal symmetry~\cite{Hirano:2017zox}.
The situation is similar if we include higher-order derivatives of $\phi$ in the action.

Second, let us comment on the relation between our extended mimetic gravity models and quadratic/cubic DHOST theories.
For a generic choice of the function~$F$ in the seed action~\eqref{genac}, the resultant mimetic action has terms of quartic or higher order in $\na_\mu\na_\nu\phi$, which cannot be reached from the quadratic/cubic DHOST class via any disformal transformation.
However, as shown in Appendix~\ref{app:cubicF}, if $F$ is of at most cubic order in $\na_\mu\na_\nu\phi$, the corresponding mimetic model falls into the quadratic/cubic DHOST class.

The third comment is on an alternative formulation of mimetic gravity.
It was pointed out in Ref.~\cite{Golovnev:2013jxa} that imposing the constraint~$X=-1$ (the {\it mimetic constraint}), as implied by Eq.~\eqref{mimcon}, leads to a theory which is equivalent to the one obtained via the noninvertible transformation~\eqref{mimtrnsf} (see also Refs.~\cite{Barvinsky:2013mea,Hammer:2015pcx}).
Note that Eq.~\eqref{mimtrnsf} reads $\ti{g}_\mn=g_\mn$ when one sets $X=-1$, which means that imposing the constraint~$X=-1$ after performing the transformation~\eqref{mimtrnsf} generates the same theory as the one obtained by imposing $\tilde{X}=-1$ from the beginning in the original frame.
In the Lagrangian formalism, the equivalence between the two formulations can directly be verified by comparing the equations of motion (EOMs).
In contrast, in the language of the Hamiltonian analysis in \S \ref{ssec:mimHam}, one could regard the mimetic constraint as a gauge condition that completely fixes the conformal gauge DOF.
Thus, we could safely impose the constraint~$X=-1$ from the beginning, which would significantly simplify the analysis.
For related arguments on eliminating ghost DOFs by constraints, see Ref.~\cite{Chen:2012au}.

Finally, we discuss the issue of the unitary gauge.
In the previous section, we did not fix the coordinate system and performed the Hamiltonian analysis maintaining general covariance.
However, in many cases it is convenient to impose the unitary gauge $\phi=t$.
Note that this gauge choice is compatible with the mimetic constraint, which ensures that $\nabla_\mu\phi$ is timelike.
Let us see that the number of DOFs of mimetic gravity does not change due to the unitary gauge fixing.
Under the unitary gauge with the mimetic constraint, one has $N=1$, $A_\ast=1$, and $V_\ast=0$.
The action of mimetic gravity~\eqref{mimac0} then reduces to
	\be
	S_{\rm unitary}=\int dtd^3x\sqrt{\ga}L(\ga_{ij},R_{ij},t;K_{ij};D_i). \label{uniac}
	\ee
The canonical variables are $(N^i,\ga_{ij};\pi_i,\pi^{ij})$, which form an 18-dimensional phase space.
If the Lagrangian~$L$ is nondegenerate, i.e., $\det \bra{\pa^2L/\pa K_{ij}\pa K_{kl}}\ne 0$, we obtain the primary constraints $\pi_i\approx 0$, and then they lead to the momentum constraints~$\mH_i\approx 0$, with no further constraints.
All these six constraints are first class and they reduce the phase-space dimension to yield
	\be
	\fr{1}{2}\bra{18-6\times 2}=3~{\rm DOFs}.
	\ee
This is consistent with the analysis without unitary gauge fixing.
On the contrary, if $L$ is degenerate, one obtains additional constraints on $\pi^{ij}$.
This is different from the case considered in the Hamiltonian analysis in \S \ref{ssec:mimHam}, as the degeneracy of $L$ in Eq.~\eqref{uniac} implies that $L_{\rm M}$ in Eq.~\eqref{mimac0} is also degenerate.
Thus, we see that the two analyses (with or without unitary gauge fixing) give consistent results in the present context of mimetic theories.

\section{Cosmological perturbations}\label{sec:pert}

The large generalization of mimetic gravity we have obtained has 3 DOFs and hence is free from obvious instabilities of Ostrogradsky ghosts.
This does not mean, however, that general mimetic theories are phenomenologically viable.
To see this point, let us analyze perturbations around the Friedmann-Lema\^itre-Robertson-Walker (FLRW) background in the mimetic gravity models.

According to the last two remarks in \S \ref{ssec:remark}, we may take safely the unitary gauge to write $\phi=t$ and impose the constraint~$X=-1$ in the action~\eqref{genac}, with which the calculation is simplified drastically.
As a consequence of $\phi=t$ and $X=-1$, any function of $(\phi,X)$ can be regarded as a function of $t$ only.
We also have $N=1$ since $X$ is written in terms of $N$ as $X=-1/N^2$ in the unitary gauge.
Therefore, the ADM form~\eqref{curvac} of the first two terms in the action~\eqref{genac} reduces to
	\be
	\int d^4x\sqrt{-g}\bra{f_2\mR+f_3\mG^\mn \na_\mu\na_\nu\phi}=\int dtd^3x\sqrt{\ga}\brb{\bra{f_2-\fr{1}{2}\dot{f}_3}R+\bra{f_2+\fr{1}{2}\dot{f}_3}\bra{K_{ij}^2-K^2}-2\dot{f}_2K}, \label{uniaccurv}
	\ee
where a dot stands for differentiation with respect to $t$.
As for the third term in Eq.~\eqref{genac}, its ADM representation~\eqref{scaac} is now written in terms of scalar quantities composed of $\ga_{ij}$ and $K_{ij}$, i.e., it can be expressed as a function of $\mK_n\equiv K^{i_1}_{i_2}K^{i_2}_{i_3}\cdots K^{i_n}_{i_1}$ ($n=1, 2, \cdots, \ell$):
	\be
	\int d^4x\sqrt{-g}F(g_\mn,\phi,\na_\mu\phi,\na_\mu \na_\nu\phi)=\int dtd^3x\sqrt{\ga}\hat{\mF}(t,K,\mK_2,\mK_3,\cdots,\mK_\ell), \label{uniacF}
	\ee
where note that ${\cal K}_1=K$.
Combining Eqs.~\eqref{uniaccurv} and \eqref{uniacF}, we obtain the following action for the mimetic counterpart of the theory~\eqref{genac}:
	\begin{align}
	S&=\int dtd^3x\sqrt{\ga}\brb{\bra{f_2-\fr{1}{2}\dot{f}_3}R+\mF(t,K,\mK_2,\mK_3,\cdots,\mK_\ell)}, \label{mimac2} \\
	\mF&\equiv \hat{\mF}+\bra{f_2+\fr{1}{2}\dot{f}_3}\bra{\mK_2-K^2}-2\dot{f}_2K.
	\end{align}
It is useful to define the first and second derivatives of $\mF$ as
	\be
	\mF_n\equiv \fr{\pa \mF}{\pa \mK_n},~~~\mF_{mn}\equiv \fr{\pa^2\mF}{\pa \mK_m \pa \mK_n},
	\ee
respectively.

Now we substitute the following metric ansatz to the action~\eqref{mimac2},
	\be
	N=1,~~~N_i=\pa_i\chi,~~~\ga_{ij}=a^2(t)e^{2\zeta}\left(e^h\right)_{ij}
	=a^2e^{2\zeta}\left(\delta_{ij}+h_{ij}+\frac{1}{2}h_{ik}h_{jk}+\cdots\right),
	\ee
where $\chi$ and $\zeta$ are scalar perturbations and $h_{ij}$ denotes a transverse-traceless tensor perturbation.
The background EOM is given by
	\be
	\dot{{\cal P}}+3H\mP-\mF=0,~~~\mP\equiv \sum_{n=1}^\ell n\,H^{n-1}\mF_n, \label{bgEOM}
	\ee
where $H\equiv \dot{a}/a$ is the Hubble parameter and $\mF_n$ are evaluated at the background, $\mK_n=3H^n$.
This equation will be used to simplify the expressions of the quadratic actions for the tensor and scalar perturbations.

The quadratic action for the tensor perturbation~$h_{ij}$ is given by
	\be
	S_{\rm T}^{(2)}=\int dtd^3x\fr{a^3}{4}\brb{\mB\dot{h}_{ij}^2-\mE\fr{(\pa_kh_{ij})^2}{a^2}}, \label{qacT}
	\ee
where
	\be
	\mB\equiv \sum_{n=2}^\ell \fr{n(n-1)}{2}H^{n-2}\mF_n,~~~\mE=f_2-\fr{1}{2}\dot{f}_3.
	\ee
The tensor perturbations are stable provided that $\mB>0$ and $\mE> 0$.

The quadratic action for the scalar perturbations~$\zeta$ and $\chi$ is
	\be
	S_{\rm S}^{(2)}=\int dtd^3xa^3\brb{\fr{3}{2}(3\mA+2\mB)\dot{\zeta}^2+2\mE\fr{(\pa_k\zeta)^2}{a^2}+\fr{1}{2}(\mA+2\mB)\bra{\fr{\pa^2\chi}{a^2}}^2-(3\mA+2\mB)\dot{\zeta}\fr{\pa^2\chi}{a^2}}, \label{qacS}
	\ee
where we have used the background EOM~\eqref{bgEOM} and defined
	\be
	\mA\equiv \sum_{m=1}^\ell \sum_{n=1}^\ell m\,n\,H^{m+n-2}\mF_{mn}.
	\ee
The EOM for $\chi$ can be solved to give
	\be
	\fr{\pa^2\chi}{a^2}=\fr{3\mA+2\mB}{\mA+2\mB}\dot{\zeta}, \label{solX}
	\ee
where we have assumed that $\mA+2\mB\ne 0$.
Substituting Eq.~\eqref{solX} to the action~\eqref{qacS}, we obtain
	\be
	S_{\rm S}^{(2)}=2\int dtd^3x\,a^3\brb{\fr{\mB(3\mA+2\mB)}{\mA+2\mB}\dot{\zeta}^2+\mE\fr{(\pa_k\zeta)^2}{a^2}}. \label{qacS2}
	\ee
Written in this form, one notices that the stability condition for the tensor perturbations, $\mE> 0$, is not compatible with the stability for the scalar perturbation, $\mE< 0$.
This indicates that either of the tensor or scalar perturbations exhibits gradient instabilities, even if one circumvents ghosts by choosing the coefficients in front of the time derivative terms in Eqs.~\eqref{qacT} and \eqref{qacS2} to be positive.
This result generalizes what was found in Refs.~\cite{Ijjas:2016pad,Firouzjahi:2017txv,Hirano:2017zox}, and we have thus established that all the mimetic gravity models with 3 DOFs obtained so far are plagued with ghost/gradient instabilities on a cosmological background (except for the special case of strongly-coupled scalar perturbations mentioned below).
To circumvent this problem from the perspective of effective field theory, one must design the models so that the time scale of the instability is longer than the cosmological time scale.
It is worth noting that the operators of the form~$\sim RK$ introduced in Refs.~\cite{Hirano:2017zox,Zheng:2017qfs} to resolve this problem are not fully satisfactory, because now it is clear from the covariant analysis of the present paper that such operators give rise to unwanted extra DOFs on a general background.

One would notice that if $\mB(\mA+2\mB)(3\mA+2\mB)=0$ then the scalar perturbations appear to be nondynamical.
This is indeed the case in the mimetic Horndeski theories~\cite{Arroja:2015wpa} where $\mA+2\mB=0$.
The situation is the same even if we start from GLPV theories.
Since such a choice of $\mA$ and $\mB$ does not change the number of DOFs, the seemingly nondynamical scalar mode should originate from strong coupling of perturbations.

A caveat should be added here.
In the case of $\mA+2\mB=0$, it is important to take into account the presence of matter fields other than $\phi$ to discuss the viability of mimetic cosmology.
Let us add to the seed Lagrangian another scalar field~$\psi$ whose Lagrangian is of the form
	\begin{align}
	L_\psi = P(\psi, Y),~~~Y\equiv -\frac{1}{2}g^{\mu\nu}\pa_\mu\psi\pa_\nu\psi.
	\end{align}
This field can also be regarded as a perfect fluid.
We split $\psi$ into the background part $\psi(t)$ and the perturbation~$\delta\psi$, and then expand the mimetic action to second order in perturbations.
In the case of $\mA+2\mB=0$ we obtain
	\begin{align}
	S_{\rm S}^{(2)}=\int dtd^3x a^3&\biggl[ -6{\cal B}\dot\zeta^2+2{\cal E}\frac{(\pa_k\zeta)^2}{a^2}+4\mB\dot\zeta \frac{\pa^2\chi}{a^2} \nonumber \\
	&+\frac{1}{2}(P_Y+2YP_{YY})\dot{\delta\psi}^2 -\frac{1}{2}P_Y(\pa_k\delta\psi)^2+\dot{\psi} P_Y \delta\psi \frac{\pa^2\chi}{a^2}+\cdots \biggr], \label{acfluid}
	\end{align}
where the ellipses represent the terms that are not relevant to the present argument.
Now the EOM for $\chi$ is given by
	\begin{align}
	4\mB\dot\zeta +\dot\psi P_Y\delta\psi = 0.
	\end{align}
Substituting this back into Eq.~\eqref{acfluid}, one can remove $\delta\psi$ as well as $\chi$ from the action.
The reduced action for $\zeta$ contains the term
	\begin{align}
	4a^3 \mB^2\left(\frac{P_Y+2YP_{YY}}{YP_Y^2}\right)\ddot\zeta^2, \label{ghosty}
	\end{align}
showing that the system has two scalar DOFs (one from $\phi$ and one from the additional matter field).
This is the reason why the scalar perturbations revive and become dynamical in mimetic Horndeski gravity in the presence of matter~\cite{Arroja:2017msd}.
It is more important to note that one of the two scalar DOFs is a ghost, as is clear from Eq.~\eqref{ghosty}.
A detailed comparison of this result with the recent work~\cite{Arroja:2017msd} is not straightforward and is beyond the scope of the present paper, because the authors of Ref.~\cite{Arroja:2017msd} investigate the EOMs and the solution in the Newtonian gauge assuming adiabatic initial conditions, while we focus on the action in the uniform $\phi$ gauge.

\section{Conclusions}\label{sec:conc}

Given that one can go beyond the Horndeski theory, i.e., the most general scalar-tensor theory with second-order Euler-Lagrange equations, by relaxing the assumptions to allow for higher-order Euler-Lagrange equations with degenerate kinetic matrix, it is intriguing to explore general healthy scalar-tensor theories with 3 DOFs.
As such ``beyond Horndeski'' theories, the broadest class known so far is the quadratic/cubic DHOST, which cannot be further extended by disformal transformation.
In the present paper, we have demonstrated that the seed action~\eqref{genac}, which is nondegenerate in general, can be transformed to give a degenerate theory through the noninvertible conformal transformation~\eqref{mimtrnsf}.
The unwanted extra DOF is eliminated by the local conformal symmetry associated with the noninvertibility of the transformation, leaving only 3 DOFs, as implied in Ref.~\cite{Liu:2017puc}.
We have shown this explicitly by means of Hamiltonian analysis.
The resultant degenerate scalar-tensor theories obtained thus are novel in the sense that they are related to none of the known healthy theories via disformal transformation, and can be thought of as an extension of mimetic gravity since the original mimetic theory is generated from the Einstein-Hilbert action through the same noninvertible conformal transformation.
It should be emphasized that not all nondegenerate scalar-tensor theories can be transformed to mimetic gravity with 3 DOFs.
Rather, we have specified the possible form of the seed action as Eq.~\eqref{genac}.
As far as we have investigated, any deviation from this seed leads to unwanted extra DOFs after the noninvertible transformation.

We have also studied cosmological perturbations in our extended mimetic gravity, and found that either of tensor or scalar perturbations is plagued with gradient instabilities in general, except for the special case where the scalar perturbations would be strongly coupled, or otherwise ghost instabilities appear.
In the strongly-coupled case, inclusion of matter fields other than the scalar field renders the scalar perturbations dynamical and unstable.

In spite of this flaw, the idea of constructing degenerate field theories from nondegenerate higher derivative theories by performing a noninvertible transformation is interesting itself and worth pursuing.
In the same way as the present case of mimetic theories, a noninvertible transformation generically introduces a symmetry in a given theory, which could eliminate unwanted DOFs that the theory originally has.
This approach may give us some clue to go beyond known healthy theories of gravity.


\acknowledgements{
This work was supported in part by JSPS Research Fellowships for Young Scientists No.~17J06778 (K.T.), 
the JSPS Grants-in-Aid for Scientific Research Nos.~16H01102 and 16K17707,
MEXT-Supported Program for the Strategic Research Foundation at Private Universities, 2014-2017 (S1411024), 
and MEXT KAKENHI Grant Nos.~15H05888 and 17H06359 (T.K.).
}


\appendix

\section{Noninvertible disformal transformations}\label{app:disf}

Here, we show that any noninvertible disformal transformation can essentially be reduced to the simplest form of Eq.~\eqref{mimtrnsf}.
Let us consider a disformal transformation
	\be
	\ti{g}_\mn=A(\phi,X)g_\mn+B(\phi,X)\pa_\mu\phi\pa_\nu\phi, \label{disf}
	\ee
with $B\ne 0$.
This transformation is invertible [namely, Eq.~\eqref{disf} is uniquely solvable for $g_\mn$] if the functions~$A$ and $B$ satisfy $A(A-XA_X-X^2B_X)\ne 0$~\cite{Bekenstein:1992pj,Zumalacarregui:2013pma,Deruelle:2014zza,Takahashi:2017zgr}, while it is noninvertible if
	\be
	A-XA_X-X^2B_X=0,~~~A(A+XB)\ne 0, \label{noninvcond}
	\ee
where the latter condition guarantees the existence of the inverse matrix~$\ti{g}^\mn$~\cite{Bekenstein:1992pj}.
We are interested in the noninvertible disformal transformations.
Equation~\eqref{noninvcond} is equivalent to
	\be
	B=-\fr{A}{X}-f(\phi),
	\ee
with $f(\phi)$ being some nonzero function of $\phi$.
As in Eq.~\eqref{mimcon}, the scalar kinetic term in the new frame, $\ti{X}$, is constrained as $\ti{X}=-1/f$.
Note that $A$ is proportional to $X$ if there is no disformal part~$B$, i.e., in the case of noninvertible {\it conformal} transformation.

Now let us consider another disformal transformation
	\be
	g_\mn=\bar{A}(\phi,\bar{X})\bar{g}_\mn+\bar{B}(\phi,\bar{X})\pa_\mu\phi\pa_\nu\phi. \label{disf2}
	\ee
Suppose that a theory~$\ti{S}[\ti{g}_\mn,\phi]$ is mapped to another theory of the form~$S[g_\mn,\phi]$ by the noninvertible disformal transformation~\eqref{disf}, and then to $\bar{S}[\bar{g}_\mn,\phi]$ by the transformation~\eqref{disf2}.
We choose the functions~$\bar{A}$ and $\bar{B}$ in Eq.~\eqref{disf2} so that the composition of the transformations~\eqref{disf} and \eqref{disf2} reduces to a noninvertible conformal transformation.
In doing so we require that the transformation~\eqref{disf2} is invertible.
The following choice of $\bar{A}$ and $\bar{B}$ satisfies these requirements:
	\be
	\bar{A}(\phi,\bar{X})\equiv \bar{Q}-\bar{X}\bar{B}(\phi,\bar{X}),~~~\bar{B}(\phi,\bar{X})\equiv -\fr{B(\phi,\bar{X}/\bar{Q})}{A(\phi,\bar{X}/\bar{Q})}. \label{dis2con}
	\ee
To ensure the invertibility of the transformation~\eqref{disf2}, $\bar{Q}$ must not be of the form~$\bar{Q}=q(\phi)\bar{X}$ [with arbitrary $q(\phi)$], but otherwise it is an arbitrary function of $(\phi,\bar{X})$.
For the choice~\eqref{dis2con}, the relation between $\ti{g}_\mn$ and $\bar{g}_\mn$ is found to be
	\be
	\ti{g}_\mn=-f(\phi)\bar{X}\bar{g}_\mn. \label{disf3}
	\ee
We see that the disformal part has been eliminated.
It should be noted that Eq.~\eqref{disf3} is independent of the function~$\bar{Q}$.
Once written in this form, the function~$f(\phi)$ can be absorbed into the redefinition of the scalar field in the following way:
Introducing a new scalar field~$\hat{\phi}$ so that $d\hat{\phi}/d\phi=f(\phi)^{1/2}$, Eq.~\eqref{disf3} is rewritten as
	\be
	\ti{g}_\mn=-\bra{\bar{g}^{\al\beta}\pa_\al\hat{\phi}\pa_\beta\hat{\phi}}\bar{g}_\mn, \label{noninvconf}
	\ee
which has the same form as Eq.~\eqref{mimtrnsf}.
Thus, instead of $S[g_\mn,\phi]$ which is obtained from $\ti{S}[\ti{g}_\mn,\phi]$ by the noninvertible disformal transformation~\eqref{disf}, we may consider $\bar{S}[\bar{g}_\mn,\phi(\hat{\phi})]$ obtained by the noninvertible conformal transformation~\eqref{noninvconf}.


\section{Noether identities}\label{app:Noether}

In this appendix, we summarize the Noether identities with respect to the spatial diffeomorphism and the conformal symmetry of the action~$S_{\rm M}[N,\ga_{ij},\phi, A_\ast,B_{ij}]$ given in Eq.~\eqref{S0}.
These identities are used to derive some of the equations in the main text.
Note that $B_{ij}$ replaces $V_{ij}$ and therefore transforms in the same way as $V_{ij}$.

\subsection{Spatial diffeomorphism}
Under an infinitesimal spatial diffeomorphism~$x^i\to x^i+\e^i(x^j)$, $N$, $\ga_{ij}$, $\phi$, $A_\ast$, and $B_{ij}$ transform as
	\begin{align}
	&\Delta_\e N=-\e^iD_iN,~~~\Delta_\e \ga_{ij}=-2D_{(i}\e_{j)},~~~\Delta_\e \phi=-\e^iD_i\phi, \\
	&\Delta_\e A_\ast=-\e^iD_i A_\ast,~~~\Delta_\e B_{ij}=-\e^kD_kB_{ij}-2B_{k(i}D_{j)}\e^k,
	\end{align}
respectively.
Since $S_{\rm M}$ is invariant under this transformation, we have
	\be
	0=\Delta_\e S_{\rm M}=-\int dtd^3x\brb{\fr{\delta S_{\rm M}}{\delta N}\e^iD_iN+2\fr{\delta S_{\rm M}}{\delta \ga_{ij}}D_{i}\e_{j}+\fr{\delta S_{\rm M}}{\delta \phi}\e^iD_i\phi+\fr{\delta S_{\rm M}}{\delta A_\ast}\e^iD_i A_\ast+\fr{\delta S_{\rm M}}{\delta B_{ij}}\bra{\e^kD_kB_{ij}+2B_{ki}D_{j}\e^k}}.
	\ee
Integrating by parts, we obtain the following relation among the variations of $S_{\rm M}$:
	\be
	\sqrt{\ga}L_{\rm M}D_iN-2\sqrt{\ga}\ga_{ij}D_k\bra{\fr{1}{\sqrt{\ga}}\fr{\delta S_{\rm M}}{\delta \ga_{jk}}}+\fr{\delta S_{\rm M}}{\delta \phi}D_i\phi+\fr{\delta S_{\rm M}}{\delta A_\ast}D_i A_\ast+\fr{\delta S_{\rm M}}{\delta B_{jk}}D_iB_{jk}-2\sqrt{\ga}D_j\bra{\fr{B_{ik}}{\sqrt{\ga}}\fr{\delta S_{\rm M}}{\delta B_{jk}}}=0. \label{3diffNoether}
	\ee

\subsection{Conformal symmetry}
Similarly, under an infinitesimal conformal transformation~$\Delta_\e g_\mn=\e(x^\la)g_\mn$, we have the following transformation law:
	\begin{align}
	&\Delta_\e N=\fr{\e}{2}N,~~~\Delta_\e \ga_{ij}=\e \ga_{ij},~~~\Delta_\e \phi=0, \\
	&\Delta_\e A_\ast=-\fr{\e}{2} A_\ast,~~~\Delta_\e B_{ij}=\fr{\e}{2}B_{ij}-\fr{D^k\phi}{2 A_\ast}\ga_{ij}D_k\e.
	\end{align}
These expressions can be obtained by replacing $(-X)$ with $1+\e$ in Eqs.~\eqref{mimN} and \eqref{mimVij}.
In the same way as in the case of spatial diffeomorphism, we obtain
	\be
	N\sqrt{\ga}L_{\rm M}=-\ga_{ij}\brb{2\fr{\delta S_{\rm M}}{\delta \ga_{ij}}+\sqrt{\ga}D_k\bra{\fr{D^k\phi}{\sqrt{\ga} A_\ast}\fr{\delta S_{\rm M}}{\delta B_{ij}}}}+ A_\ast\fr{\delta S_{\rm M}}{\delta A_\ast}-B_{ij}\fr{\delta S_{\rm M}}{\delta B_{ij}}. \label{conNoether}
	\ee


\section{Horndeski's conformally invariant scalar-tensor theory}\label{app:Horndeski}

In Ref.~\cite{Horndeski:2017rtl}, Horndeski specified all the conformally invariant scalar-tensor theories that are {\it flat space compatible}, i.e., such that one can take the limit~$g_\mn\to \eta_\mn$ and $\phi\to{\rm constant}$.
These theories are described by the following action:
	\be
	S_{\rm CI}=\int d^4x\sqrt{-g}\bra{L_{\rm 2C}+L_{\rm 3C}+L_{\rm 4C}+L_{\rm UC}}, \label{CIST}
	\ee
where
	\be
	\begin{split}
	L_{\rm 2C}&\equiv \ka_2(\phi)X^2, \\
	L_{\rm 3C}&\equiv \fr{1}{2}\ka_3(\phi)\varepsilon^{\mn\la\si}\mW^{\al\beta}{}_\mn \mW_{\al\beta\la\si}, \\
	L_{\rm 4C}&\equiv \ka_4(\phi)\mW^{\mn\la\si}\mW_{\mn\la\si}, \\
	L_{\rm UC}&\equiv \ka_{\rm U}(\phi)\brb{2X\mR-3(\Box \phi)^2-6(\na_\mu\na_\nu\phi)^2-12(\na^\mu\phi)\Box(\na_\mu\phi)},
	\end{split} \label{CIlag}
	\ee
with $\mW_{\mn\la\si}$ being the 4-dimensional Weyl tensor defined as $\mW^\mn{}_{\la\si}\equiv \mR^\mn{}_{\la\si}-2\delta^{[\mu}_{[\la}\mR^{\nu]}_{\si]}+\fr{1}{3}\delta^{\mu}_{[\la}\delta^{\nu}_{\si]}\mR$.
Note that $L_{\rm 3C}$ can be expressed equivalently as $L_{\rm 3C}=\fr{1}{2}\ka_3(\phi)\varepsilon^{\mn\la\si}\mR^{\al\beta}{}_\mn \mR_{\al\beta\la\si}$~\cite{Grumiller:2007rv}, which is nothing but the Chern-Simons term coupled with a scalar field~\cite{Jackiw:2003pm}.
One can easily check that the above action is invariant under the conformal rescaling of the metric only:~$g_{\mu\nu}\to \Omega^2g_{\mu\nu}$, $\phi\to\phi$.\footnote{Hence, the ``conformally invariant scalar-tensor theory'' is different from the theory with a ``conformally coupled scalar field'' where the action is invariant under the simultaneous transformation of the metric and the scalar field:~$g_\mn \to \Om^2g_\mn, \phi \to \Om^{-1}\phi$.}

Among the four terms in the action~$S_{\rm CI}$, $L_{\rm 2C}$ and $L_{\rm UC}$ are obtained by the noninvertible conformal transformation~\eqref{mimtrnsf} of the following theory:
	\be
	S=\int d^4x\sqrt{-\ti{g}}\brc{\ka_2(\phi)+\ka_{\rm U}(\phi)\brb{2\ti{\mR}-3\bra{\ti{\Box}\phi}^2+6\bra{\ti{\na}_\mu\ti{\na}_\nu\phi}^2}},
	\ee
which is a particular case of the general action~\eqref{genac}.
If one starts from a more general scalar-tensor theory~\eqref{genac} and performs the transformation~\eqref{mimtrnsf}, the resultant action still has the same conformal symmetry, but does not necessarily fall into the above class.
This is due to the assumption in Ref.~\cite{Horndeski:2017rtl} that the theory is flat space compatible, which plays a crucial role in narrowing down the candidate Lagrangians to a finite number as presented in Eq.~\eqref{CIlag}.

As was the case with theories obtained via the transformation~\eqref{mimtrnsf}, the theory~\eqref{CIST} generically suffers from instability in cosmological perturbations.
The theory described by $L_{\rm 2C}$ and $L_{\rm UC}$ amounts to the choice
	\be
	f_2=2\ka_{\rm U}(t),~~~f_3=0,~~~\mF=\ka_{\rm U}(t)(8\mK_2-5K^2)-4\dot{\ka}_{\rm U}(t)K+\ka_2(t)
	\ee
in Eq.~\eqref{mimac2}, for which we have $\mA=-10\ka_{\rm U}$ and $\mB=8\ka_{\rm U}$.
Therefore, the same ghost/gradient instabilities as in \S \ref{sec:pert} arise.
As for the other terms $L_{\rm 3C}$ and $L_{\rm 4C}$, the action contains higher-order derivatives of the metric, which results in Ostrogradsky ghosts.


\section{Mimetic quadratic/cubic scalar-tensor theories}\label{app:cubicF}

In this appendix, we study the particular case where the function~$F$ in the action~\eqref{genac} contains up to cubic terms of $\na_\mu\na_\nu\phi$, and consider to perform the noninvertible conformal transformation~\eqref{mimtrnsf}.
Up to cubic order, the most general function~$F$ is given by~\cite{BenAchour:2016fzp}
	\be
	F(g_\mn,\phi,\phi_\mu,\phi_\mn)=F_0+F_1\Box\phi+\sum_{i=1}^5a_iL_i^{(2)}+\sum_{j=1}^{10}b_jL_j^{(3)},
	\ee
where $F_0,F_1,a_i$, and $b_j$ are arbitrary functions of $(\phi,X)$, and we denote $\phi_\mu\equiv \na_\mu\phi$ and $\phi_\mn\equiv \na_\mu\na_\nu\phi$.
The building blocks~$L_i^{(2)}$ and $L_j^{(3)}$ are defined as
	\be
	L_1^{(2)}=\phi_\mu^\nu\phi_\nu^\mu,~~~L_2^{(2)}=\bra{\Box\phi}^2,~~~L_3^{(2)}=\bra{\Box\phi}\phi^\mu\phi_\mu^\nu\phi_\nu,~~~	L_4^{(2)}=\phi^\mu\phi_\mu^\nu\phi_\nu^\la\phi_\la,~~~L_5^{(2)}=\bra{\phi^\mu\phi_\mu^\nu\phi_\nu}^2,
	\ee
and
	\be
	\begin{split}
	&L_1^{(3)}=\bra{\Box\phi}^3,~~~L_2^{(3)}=\bra{\Box\phi}\phi_\mu^\nu\phi_\nu^\mu,~~~L_3^{(3)}=\phi_\mu^\nu\phi_\nu^\la\phi_\la^\mu,~~~L_4^{(3)}=\bra{\Box\phi}^2\phi^\mu\phi_\mu^\nu\phi_\nu,~~~\\
	&L_5^{(3)}=\bra{\Box\phi}\phi^\mu\phi_\mu^\nu\phi_\nu^\la\phi_\la,~~~L_6^{(3)}=\bra{\phi_\mu^\nu\phi_\nu^\mu}\bra{\phi^\mu\phi_\mu^\nu\phi_\nu},~~~L_7^{(3)}=\phi^\mu\phi_\mu^\nu\phi_\nu^\la\phi_\la^\si\phi_\si, \\
	&L_8^{(3)}=\bra{\phi^\mu\phi_\mu^\nu\phi_\nu^\la\phi_\la}\bra{\phi^\si\phi_\si^\rho\phi_\rho},~~~L_9^{(3)}=\bra{\Box\phi}\bra{\phi^\mu\phi_\mu^\nu\phi_\nu}^2,~~~L_{10}^{(3)}=\bra{\phi^\mu\phi_\mu^\nu\phi_\nu}^3.
	\end{split}
	\ee
If the coefficients~$a_i$ and $b_j$, as well as~$f_2$ and $f_3$ in the original action~\eqref{genac} are chosen appropriately, the theory reduces to quadratic/cubic DHOST~\cite{Langlois:2015cwa,Crisostomi:2016czh,BenAchour:2016fzp}.

Now we perform the transformation~\eqref{mimtrnsf}.
It should be noted that the old-frame coefficients, which are originally functions of $(\phi,\ti{X})$, are now interpreted as functions only of $\phi$ because $\ti{X}=-1$ [see Eq.~\eqref{mimcon}].
Moreover, with the aid of this constraint on $\ti{X}$, the term with the Einstein tensor in the action~\eqref{genac} can be integrated by parts to give
	\begin{align}
	\int d^4x\sqrt{-\ti{g}}\ti{f}_3\ti{\mG}^\mn\ti{\na}_\mu\ti{\na}_\nu\phi=
	\int d^4x\sqrt{-\ti{g}}\brc{-\ti{f}_{3\phi}\brb{\fr{1}{2}\ti{\mR}+\bra{\ti{\Box}\phi}^2-\bra{\ti{\na}_\mu\ti{\na}_\nu\phi}^2}
		+\ti{f}_{3\phi\phi\phi}},
	\end{align}
from which we see that the contribution of the $\ti{f}_3$ term can be absorbed into $\ti{f}_2,\ti{a}_1,\ti{a}_2$, and $\ti{F}_0$.
Thus, we set $\ti{f}_3=0$ from the beginning without loss of generality.
The new-frame action also belongs to the same class of theories as the original one, and the functions~$f_2,a_i,b_j$ are written in terms the old-frame functions~$\ti{f}_2,\ti{a}_i,\ti{b}_j$ as follows:
	\be
	f_2=-X\ti{f}_2,
	\ee
	\be
	a_1=\ti{a}_1,~~~a_2=\ti{a}_2,~~~a_3=\fr{2}{X}\bra{\ti{a}_1+2\ti{a}_2},~~~a_4=-\fr{2}{X}\bra{\ti{a}_1+3\ti{f}_2},~~~a_5=\fr{2}{X^2}\bra{\ti{a}_1+2\ti{a}_2},
	\ee
	\be
	\begin{split}
	&b_1=-\fr{1}{X}\ti{b}_1,~~~b_2=-\fr{1}{X}\ti{b}_2,~~~b_3=-\fr{1}{X}\ti{b}_3,~~~b_4=-\fr{2}{X^2}\bra{3\ti{b}_1+\ti{b}_2},~~~b_5=\fr{2}{X^2}\ti{b}_2,~~~b_6=-\fr{1}{X^2}\bra{2\ti{b}_2+3\ti{b}_3}, \\
	&b_7=\fr{3}{X^2}\ti{b}_3,~~~b_8=\fr{1}{X^3}\bra{4\ti{b}_2+3\ti{b}_3},~~~b_9=-\fr{3}{X^3}\bra{4\ti{b}_1+2\ti{b}_2+\ti{b}_3},~~~b_{10}=-\fr{2}{X^4}\bra{4\ti{b}_1+2\ti{b}_2+\ti{b}_3}, \\
	\end{split}
	\ee
while the functions~$F_0$ and $F_1$ are not important.
Note that there is no contribution of $\ti{a}_3,\ti{a}_4,\ti{a}_5$, and $\ti{b}_4,\cdots,\ti{b}_{10}$ because the corresponding building blocks contain $\ti{\na}_\mu\phi\ti{\na}^\mu\ti{\na}_\nu\phi$, which is mapped to zero by the transformation~\eqref{mimtrnsf}.
Since the resultant mimetic theory has at most three DOFs (see \S \ref{ssec:mimHam}), it must be a DHOST theory.
Indeed, one can show that this theory is represented as a combination of ${}^2$N-III and ${}^3$M-I theories in the terminology of Ref.~\cite{BenAchour:2016fzp}, and hence is degenerate.


\bibliography{mimetic}

\end{document}